\documentstyle[preprint,aps,epsf]{revtex}

\newcommand{\ber}{\begin{eqnarray}}
\newcommand{\eer}{\end{eqnarray}}
\begin{document} 

\begin{titlepage} 
\flushright{LBNL-40219}
\begin{center}
\Large {\bf Energy loss effects on charm and bottom production
in high-energy heavy-ion collisions
}
\end{center}
\begin{center}
{Ziwei Lin$^a$, Ramona Vogt$^{a,b}$ and Xin-Nian Wang$^a$
}\\[2ex]
$^a$Nuclear Science Division, LBNL, Berkeley, CA 94720 \\
$^b$Physics Department, University of California at Davis, Davis, CA 95616
\\ [2em] 
\end{center}

\begin{abstract}
We study the effect of energy loss on charm and bottom quarks in high-energy
heavy-ion collisions including longitudinal expansion and partial 
thermalization. We find that high $p_\perp$ heavy quarks are greatly 
suppressed, and consequently, high-mass dileptons from heavy quark decays are
also suppressed.  
We consider in detail the detector geometry and single lepton energy cuts 
of the PHENIX detector at the Relativistic Heavy Ion Collider (RHIC). 
Because of the longitudinal expansion, the suppressions of $ee, e\mu$ and
$\mu\mu$ pairs resulting from the heavy quark energy loss are very different 
due to the different rapidity coverages and energy cuts.  
The assumed energy loss rate, on the order of 1 GeV/fm, 
results in a large suppression on dielectrons, and 
dielectrons from heavy quark decays become comparable or even lower than the
Drell-Yan yield.
It is thus possible to probe the energy loss rate of the medium using 
dileptons from heavy quark decays.
\end{abstract}
\end{titlepage} 

\section{Introduction}

A dense parton system is expected to be formed in the early stage of
relativistic heavy-ion collisions at RHIC energies and above, due to
the onset of hard and semi-hard parton scatterings. Interactions among
the produced partons in this dense medium will most likely lead to 
partial thermalization and formation of a quark-gluon plasma. It is thus
important to study phenomenological signals of the early parton
dynamics, a crucial step towards establishing evidence
for a strongly {\it interacting} initial system and its approach to
thermal equilibrium.  Jet quenching is a very good candidate for
such a signal since a fast parton traversing dense matter must 
experience multiple scatterings  and suffer radiative energy 
loss\cite{gw_lpm,bdps,bdmps}. Taking into account multiple scatterings 
and the so-called Landau-Pomeranchuk-Midgal effect, the energy loss of 
a fast quark, $-dE/dx \simeq 3 \alpha_s <p_{\perp w}^2>/8$\cite{bdmps}, 
is found to be completely controlled by the the characteristic broadening 
of the transverse momentum squared of the fast parton, $<p_{\perp w}^2>$,
which in turn is determined by the properties of the medium.  
Therefore the energy loss of fast partons is a good probe of dense 
matter \cite{gw_jet}.

In principle, the energy loss by a parton in medium, both by 
radiative\cite{gw_lpm,bdps,bdmps} and elastic\cite{bt,tg} processes, is
independent of the quark mass in the infinite energy limit.  
At finite energies, studies show that the elastic 
energy loss has a weak mass dependence.  For example, at 10 GeV with
$\alpha_s$=0.2, $n_f=2$ and a temperature of 250 MeV, the elastic $dE/dx$ for a
massless quark and a charm quark is about $-$0.25 and $-$0.22 GeV/fm,
respectively. 
We expect that the situation for radiative energy loss is similar, though
a detailed analysis is beyond the scope of this paper.  Instead, we study the
phenomenological consequences of heavy quark energy loss with several values
of $dE/dx$ ($-$2, $-$1 and $-$0.5 GeV/fm).

Since parton energy loss results in the suppression of leading
particle productions from the corresponding jet, the particle spectra in
central $A+A$ collisions and their modification from $pp$ collisions,
either in normal events \cite{gw_jet} or photon-jet events \cite{whs}, 
can be used to study the parton energy loss. For this reason, energetic
charm and bottom mesons are particularly good leading particles
because they carry most of the charm or bottom quark's energy.
The energy loss of fast charm or bottom quarks will be directly
reflected in the suppression of large $p_T$ charm or bottom mesons.

Unfortunately, it is difficult to detect charm or bottom mesons directly
with current tracking technology because of the large number of produced 
particles in central $A+A$ collisions at RHIC. However, it was pointed out
in earlier studies \cite{vogt1,vogt2,shv} that leptons from the decay
of charm mesons will dominate the dilepton spectrum over that from
bottom decays and the Drell-Yan process in the large invariant mass 
region beyond the $J/\psi$. Since the invariant mass of the
lepton pairs from charm decays is related to the relative momentum
of the $D\bar{D}$ pair, the dilepton yields in this region could 
become an indirect measurement of the charm spectrum. Therefore, it
should be sensitive to the energy loss suffered by the charm quarks
if they propagate through dense matter. Shuryak \cite{shuryak}
recently studied the effect of the energy loss on heavy quarks in 
$A+A$ collisions and found that heavy quark pairs at large invariant mass 
are suppressed. Consequently dileptons from 
their decays are also suppressed. 
However, this study only considered heavy quarks in the central rapidity
region.  Moreover, thermal fluctuations are also
neglected so that most heavy quarks are at rest, stopped in the dense medium.  

In this paper, we re-examine the effects of heavy quark energy loss, 
including the longitudinal expansion and thermal fluctuations, which 
are important for the dilepton spectrum from heavy quark decays.
Because of the longitudinal expansion, the momentum loss in the longitudinal
direction is quite different from that in the transverse direction. 
The resulting suppression of high invariant mass dileptons 
from heavy quark decays is then very sensitive to the phase space
restrictions imposed by the detector design, {\it e.g.}, PHENIX at RHIC. 
Heavy quarks cannot be at rest 
in a thermal environment. In the most
extreme scenario when they are thermalized, they must have a 
thermal momentum distribution in their local frame. The
resulting heavy quark spectrum and the corresponding dilepton
spectrum will then be different from that obtained
by Shuryak \cite{shuryak}. We use a Poisson 
distribution for the number of scatterings to model the effects of
fluctuations in the number of
multiple scatterings in a finite system. In this case,
the heavy quarks have a finite probability of escaping the
system without interactions or energy loss.

This paper is organized as follows. We first explain our simple model
for energy loss in section~\ref{sec-model}. In section~\ref{sec-hvq},
we discuss the effect of energy loss on the charm and bottom quark spectra 
within our model.
In section~\ref{sec-dilepton}, we show the resulting dilepton spectra from
correlated heavy quark decays.   
To demonstrate the sensitivity to the phase space restriction, 
in section~\ref{sec-phenix} we calculate the spectra of 
$e^+e^-,e^{\pm}\mu^{\mp}$ and $\mu^+\mu^-$ pairs from correlated charm decays
within the planned acceptance of the PHENIX detector, taking into account the
detector geometry and single lepton energy cuts.
If high mass dileptons from open charm decays are suppressed by the energy
loss, dileptons from bottom decays can become important, 
especially after imposing the single lepton energy cuts, 
because these decays can generate more energetic leptons 
due to the larger bottom quark mass.  
Therefore we study the energy loss effects on both charm and bottom quarks in
this paper. 
To study the sensitivity to the longitudinal expansion pattern of the dense
matter, we compare the results from our model with those from a static fireball
model where the longitudinal expansion and the thermal effect are neglected so
that the heavy quarks energy loss is isotropic.  
The energy loss rate, $dE/dx$, is $-$2 GeV/fm if not specified otherwise.  
We also  study the sensitivity of single 
lepton and dilepton production to the magnitude of the energy loss by comparing
results with different values of $dE/dx$.
In section~\ref{sec-single} we calculate the single $e$ and $\mu$ 
spectra from charm and bottom decays within the PHENIX acceptance.  
We summarize in section~\ref{sec-summary}.

\section{The Model}
\label{sec-model}

In order to implement the energy loss of heavy quarks, we need to specify
the phase space distribution of the heavy quarks and the space-time evolution 
of the dense matter. For heavy-ion collisions at collider energies, we 
use the Bjorken model \cite{bj} to evolve the system. 
The dense matter then has a fluid velocity $v^F_z=z/t$ in the
longitudinal direction. This is essentially the fluid velocity of
free-streaming particles produced at $z=0$ and $t=0$. We neglect the 
transverse flow which sets in at a much later time. We also 
neglect the formation times of both the heavy quarks and the dense medium.
Therefore heavy quarks are produced at $z=0$, 
the same point at which the expansion begins.
Then, for any given space-time point, ($z,t$), a heavy quark
will find itself in a fluid with the same longitudinal velocity.
Therefore in the fluid rest frame, the heavy quark has momentum $(0,\vec
p_\perp)$. 
Energy loss reduces the heavy quark momentum to $(0,\vec{p_\perp}^\prime)$ in
the fluid rest frame so that the momentum of the heavy quark changes
from $(m_\perp \sinh y, \vec p_\perp)$ to $(m^\prime_\perp \sinh y, 
\vec{p_\perp}^\prime)$ in the lab frame.  
Thus a heavy quark loses its transverse momentum but retains its rapidity
because it follows the longitudinal flow.  
We will only consider fluctuations in the longitudinal momentum when a quark
loses most of its energy as we shall describe below.

Since we neglect the transverse expansion, the transverse 
area is the area of the nucleus.  The probability of heavy quark 
production as a function of the transverse radius is proportional 
to the overlap function of the two nuclei at zero impact parameter. To
simplify the calculations, we assume spherical nuclei of radius 
$r_A=r_0 A^{1/3}$.  
Since the heavy quarks have a finite probability to
escape without interaction or energy loss, we introduce the 
mean-free-path of the heavy quark in the dense matter, $\lambda$, 
in addition to the average energy loss per unit length, $dE/dx$.
For a heavy quark with a transverse path, $l_\perp$, in the medium, 
$\mu=l_\perp/\lambda$ gives the average number of scatterings.
We then generate the actual number of scatterings, $n$, from the Poisson
distribution, $P(n,\mu)=e^{-\mu} \mu^n/n!$.  
This corona effect is particularly important for heavy quarks produced at the
edge of the transverse plane of the collision.  
In the rest frame of the medium, the heavy quark then experiences momentum loss
$\Delta p = n \lambda \; dE/dx$.

When a heavy quark loses most of its momentum in the fluid rest frame,  it 
begins to thermalize with the dense medium.  
To include this thermal effect, we consider the heavy quark to be 
thermalized if its final transverse momentum after energy loss, 
$p^\prime_\perp$, is smaller than the average transverse 
momentum of thermalized heavy quarks with a temperature $T$.  
These thermalized heavy quarks are then given a random thermal momentum in the
rest frame of the fluid generated from the thermal distribution
$dN/d^3p \propto \exp {(-E/T)}$. The final momentum of the thermalized
heavy quark is obtained by transforming back from the local fluid frame with
a velocity of $z/t$ to the center-of-mass
frame of the collision. In our calculations, we assume
$\lambda=1$ fm and $T=150$ MeV. 

\section{Effects of Energy Loss on Heavy Quarks}
\label{sec-hvq}

We generate the momentum distribution of $c \bar c$ pairs from the
HIJING program\cite{hijing}.  
Initial and final state radiation effectively simulates
higher-order contributions to charm production so that the pair is no
longer azimuthally back-to-back as at leading order.
In principle, the momentum distribution of the pair can be 
calculated at next-to-leading order \cite{nlo}.  In practice, however, 
next-to-leading 
order numerical programs\cite{vogt3,ina} must
cancel singularities in order to obtain a finite result\cite{mnr}.
This is relatively easy if one integrates the cross section over 
part of the phase space.  It is, however, difficult to numerically cancel these
divergences when the 
complete kinematics of the heavy quark pair are desired.
Since we would like to use the dilepton spectra from correlated 
charm decays as an indirect measurement of charm production, it
is imperative that we have the complete information of the charm
pair, including the momentum correlation between the $c$ and $\overline c$.

In the calculation of open charm production with the HIJING program, we use the
MRS D$-^\prime$\cite{mrsd-p} parton distribution functions with $m_c=1.3$ GeV. 
We normalize the calculation so that the charm pair production cross section is
340 $\mu$b in $\sqrt{s} = 200$ GeV $pp$ collisions at RHIC.  
With an $A^{4/3}$ scaling and an 
inelastic $pp$ cross section of 40 mb, this results in an average of
9.7 charm pairs in a central Au+Au event at RHIC.  Shadowing effects on 
the nuclear parton distribution functions, 
not large for large invariant mass pairs, are not included.

\begin{figure}[h]
\setlength{\epsfxsize=\textwidth}
\setlength{\epsfysize=0.6\textheight}
\centerline{\epsffile{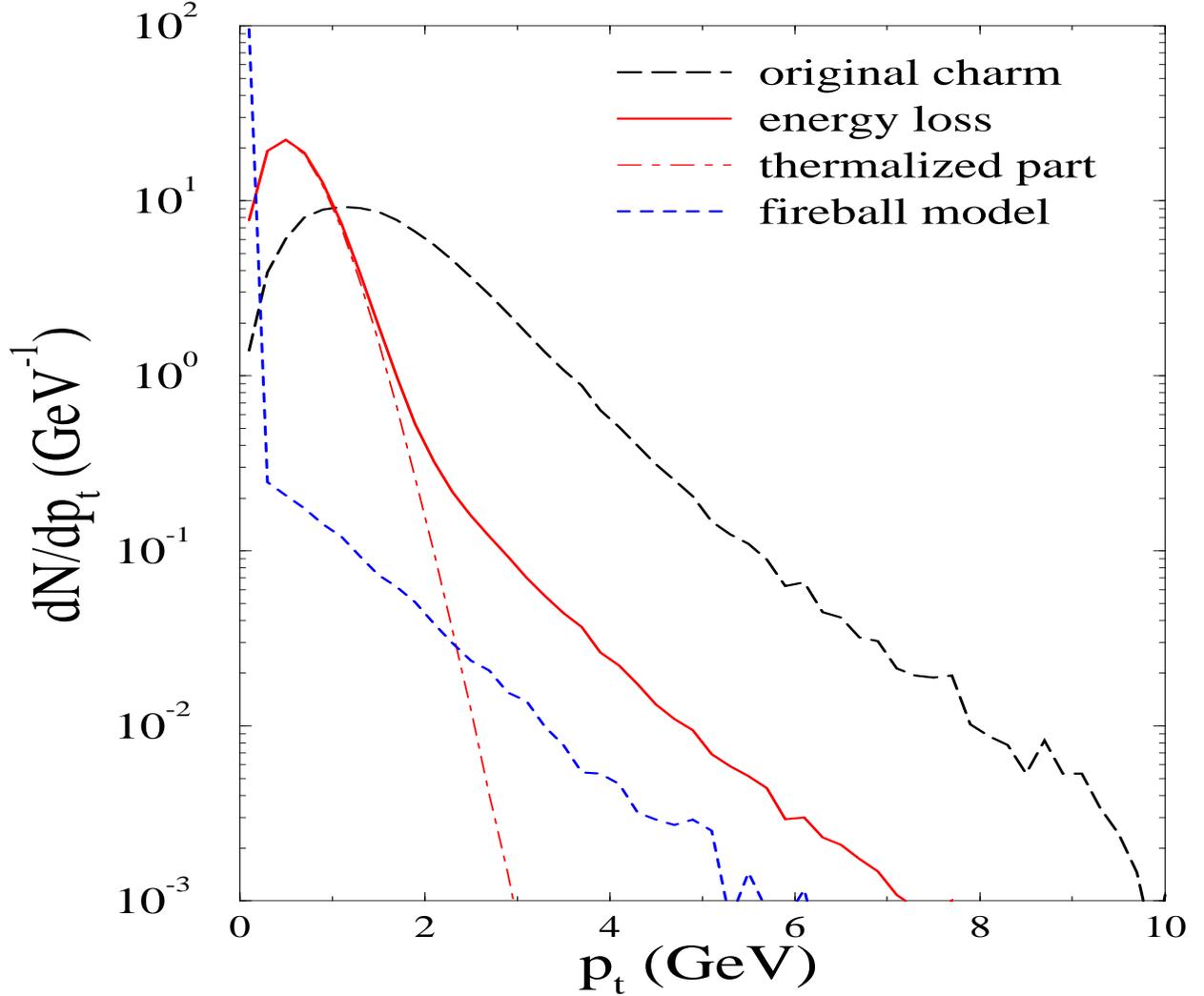}}
\vspace{1cm}
\caption{
The $p_\perp$ spectrum of charm and anti-charm quarks in a central Au+Au
collision at $\protect \sqrt s=200$ GeV.  
The long-dashed curve represents the initial production without energy loss.
The solid curve is our result with the energy loss ($dE/dx$ = $-$2 GeV/fm) 
and the dot-dashed curve represents the part from thermalized charm
quarks.  The short-dashed curve is calculated with the static fireball model.
}
\label{fig_pt_c}
\end{figure}

We now study the effect of energy loss on the charm quark spectra
in the simple model we have described. Fig.~\ref{fig_pt_c} shows 
the $p_\perp$ spectrum of charm and anti-charm quarks. 
It is normalized to one central Au+Au event at RHIC, as are all the other
figures.  
With $dE/dx=-2$ GeV/fm, $\lambda=1$ fm, and $T=150$ MeV for
thermalized charm, the spectrum (solid) is softer than the 
initial distribution (long-dashed). The yield of high $p_\perp$ charm quarks 
is suppressed by an order of magnitude.  
High $p_\perp$ charm quarks are mostly those 
with enough energy to escape the dense matter.  After losing
much of its momentum, a charm quark will become part of the thermalized
system. The $p_\perp$ distribution of these charm quarks is shown 
as the short-dashed curve in Fig.~\ref{fig_pt_c}.
In the static fireball model, where expansion 
and thermal effects are neglected, 
most of the charm quarks are stopped inside the dense matter. Therefore,
the spectrum has a peak at zero $p_\perp$. Since the corona effect
is also neglected in the fireball model, high-$p_\perp$ charm
quarks are suppressed more than in our model. This difference
is also reflected in the dilepton spectrum from charm decays.

\begin{figure}[h]
\setlength{\epsfxsize=\textwidth}
\setlength{\epsfysize=0.6\textheight}
\centerline{\epsffile{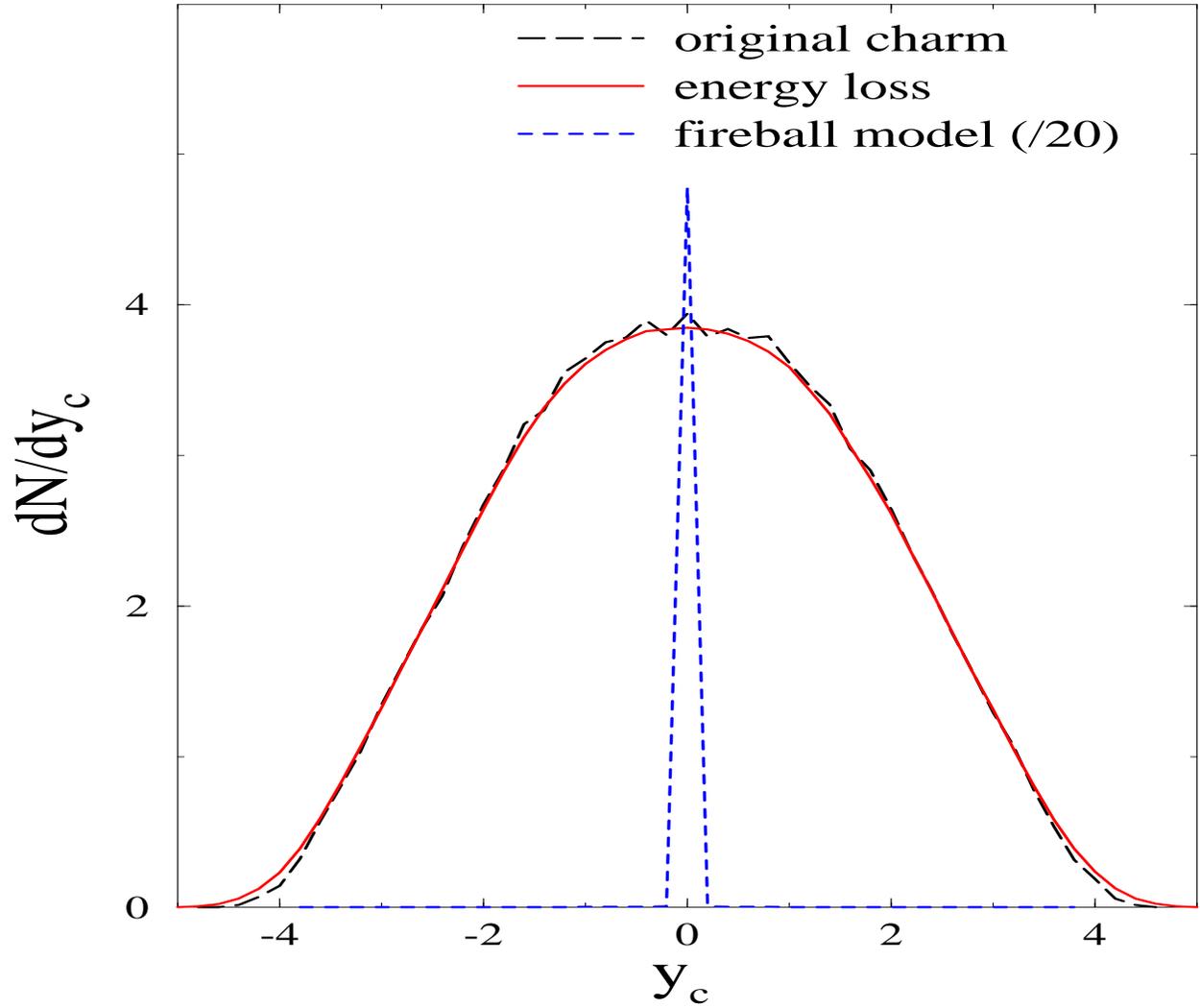}}
\vspace{1cm}
\caption{
Rapidity spectrum of charm and anti-charm quarks.  
The short-dashed curve, calculated with the static fireball model, is scaled
down by a factor of 20.
}
\label{fig_y_c}
\end{figure}

In Fig.~\ref{fig_y_c} we plot the single charm rapidity distribution
with (solid and short-dashed) and  without (long-dashed) energy loss.
Since charm quarks go with the longitudinal flow in our model, there 
is little difference between the rapidity spectrum after the energy loss and 
the initial spectrum.  
In the static fireball model (short-dashed) 
however, most charm quarks are stopped at zero rapidity since there is 
no longitudinal flow.  This sensitivity of the charm rapidity 
distribution to the longitudinal flow pattern can provide information on the
expansion of the dense matter.  
The initial rapidity distribution of charm quarks is not very
different from the gluon rapidity distribution.  
If the initial pressure causes a strong additional longitudinal expansion,
the final charm rapidity distribution will  probably become
broader, similar to other produced particles.

In order to obtain the final $D$ meson distributions, one should
convolute the charm quark distribution with a fragmentation function.
In hadronic collisions, 
one can assume a delta-function-like fragmentation function, $\delta(1-z)$,  
for charm quark fragmentation into $D$ mesons.  As a result, the $D$
meson keeps all the momentum of its charm-quark parent\cite{frag,e769}.
Figs.~\ref{fig_pt_c} and \ref{fig_y_c} then also describe the momentum 
spectrum of $D$ mesons.  
However, the fragmentation function could be different
in a dense medium than in $pp$ collisions.  
This uncertainty is beyond the scope of this paper.

We also use the HIJING program to calculate $b \bar b$ pair production.   
The bottom pair production cross section with $m_b=4.75$ GeV is 1.5 $\mu$b
in $\sqrt{s} = 200$ GeV $pp$ collisions at RHIC, extrapolating to 
0.043 $b\bar b$ pairs on average in a central Au+Au event.  
Although the energy loss experienced by 
bottom quarks may be different from that of charm
quarks, for simplicity we take the same parameters, $dE/dx, \lambda$ 
and $T$ as for the charm quark energy loss. 
High $p_\perp$ bottom quarks are suppressed by an order of magnitude, 
similar to the charm results shown in Fig.~\ref{fig_pt_c}.  
We emphasize that the energy loss rate for a heavier quark could be smaller
than that for a lighter quark.  We will study the sensitivity
of our results to the phenomenological $dE/dx$ in section~\ref{sec-phenix}.

\section{Dileptons from Heavy Quark Decays}
\label{sec-dilepton}

One can use the
dilepton spectrum from charm decays to indirectly measure charm quark 
production when a direct measurement via tracking is difficult.
Measurements of high-mass dileptons are themselves important.  
Copious thermal dilepton production\cite{thermal} was proposed as 
a signal of the formation of a fully thermally and chemically equilibrated
quark-gluon plasma. In order to obtain the yields of thermal 
dileptons, one needs to subtract the background from heavy quark decays.  
When energy loss was not included, dileptons from open charm decays 
at RHIC were shown to be about an order of magnitude higher than 
the contributions from the Drell-Yan process and bottom decays\cite{vogt2,shv},
making them the dominant background to the proposed thermal dileptons.
Energy loss changes the heavy quark momentum distribution as well as  
the resulting dilepton spectra from heavy quark decays.
Therefore, understanding the effect of energy loss on dileptons 
from heavy quark decays is also an important step towards the 
observation of thermal dilepton signals.

We computed the dilepton spectra from $D$ meson decays with a Monte Carlo code 
using JETSET7.4 \cite{jetset} for $D$ meson decays.  
The average branching ratios of $\bar c \rightarrow e X$ and 
$\bar c \rightarrow \mu X$ are taken to be 12\%. 
The lepton energy spectrum from $D$ meson semileptonic decays in
JETSET7.4 is consistent with the measurement of the MARK-III
collaboration\cite{mark3}.  
In this paper, only dileptons from correlated charm pair decays are considered,
{\it i.e.}, a single $c \bar c$ pair produces the dilepton.  
The dilepton spectrum from uncorrelated charm pair 
decays, higher and softer than that from correlated charm pair decays, 
may be removed by like-sign subtraction.  

\begin{figure}[h]
\setlength{\epsfxsize=\textwidth}
\setlength{\epsfysize=0.6\textheight}
\centerline{\epsffile{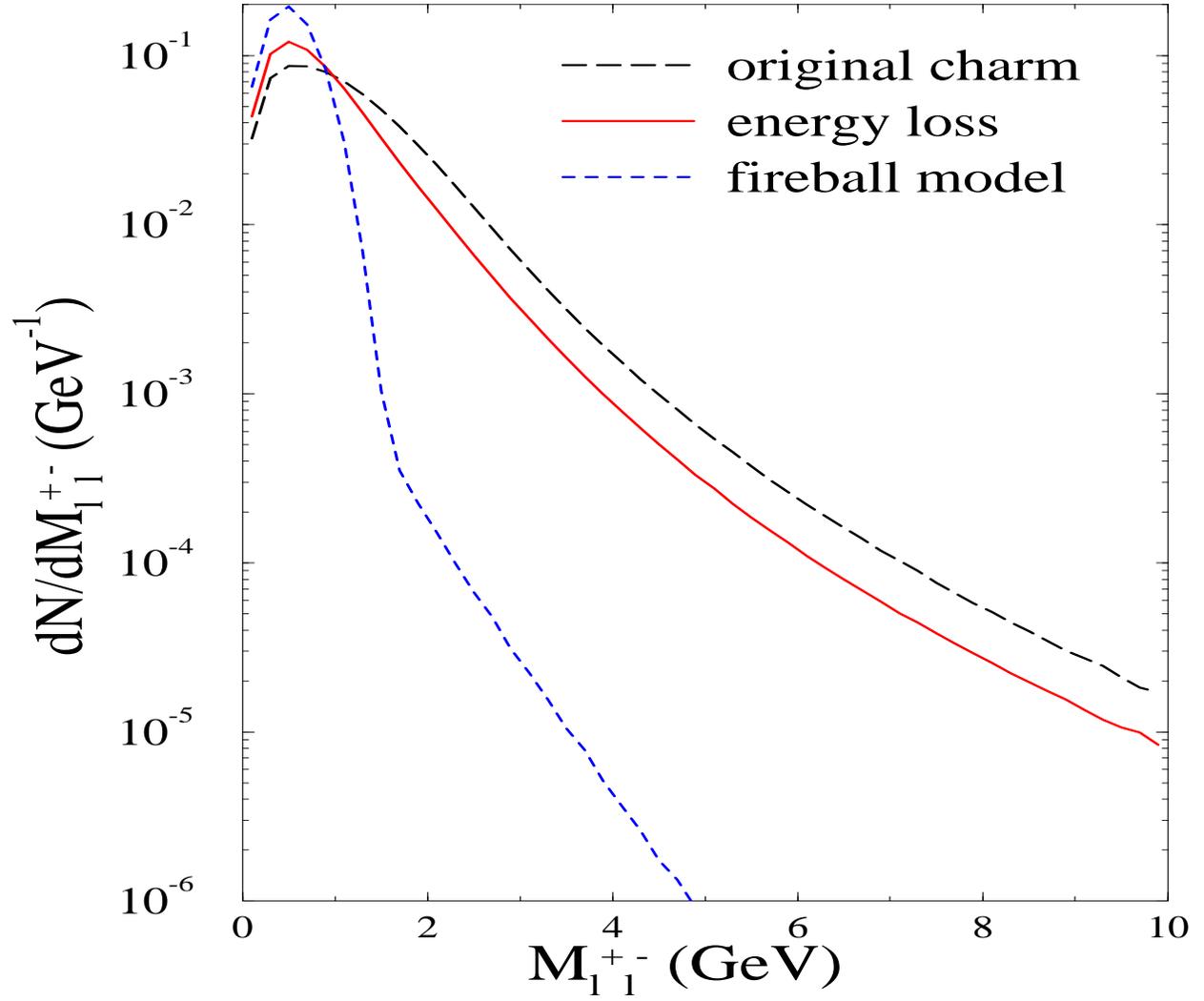}}
\vspace{1cm}
\caption{
Invariant mass spectrum of dileptons from correlated charm pair decays.  
}
\label{fig_charmll}
\end{figure}

The dilepton invariant mass and rapidity are defined as:
\ber
M_{l^+l^-}&=&\sqrt {(p^\nu_{l^+}+p^\nu_{l^-})^2} \nonumber \\
Y_{l^+l^-}&=&\tanh ^{-1} \frac {p_{l^+}^z +p_{l^-}^z} {E_{l^+} +E_{l^-}}
\, \, . \eer
In Fig.~\ref{fig_charmll}, we show the dilepton invariant mass spectrum 
from correlated charm pair decays without any phase space cuts.  
Except for the small difference between the electron and 
muon masses, this spectrum represents both dielectrons and dimuons while the 
spectrum of opposite-sign $e\mu$ is a factor of two larger.  
There is a small suppression at large invariant mass 
in our calculation with energy loss compared to the original spectrum while 
the yield from the static fireball model is reduced by about two orders of
magnitude relative to the original. The two-component nature of the
dilepton spectrum from the fireball model can be easily understood.  The
low-mass peak mainly arises from decays of $D$ mesons at rest 
in the dense medium while the high-mass tail comes 
from $D$ mesons energetic enough to escape the system.
The dilepton spectrum from the static fireball model and the resultant
suppression relative to the original spectrum are very similar to the results
from Shuryak's study\cite{shuryak} where only mid-rapidity heavy quarks are
considered. 

The lepton spectra from bottom decays are also generated from JETSET7.4.
We assume $b$ quarks fragment into $B^-, \bar B^0, \bar {B^0_s}$ 
and $\Lambda_b^0$ with the production fractions of 38\%, 38\%, 11\% 
and 13\%, respectively. Single leptons from bottom decays can be 
categorized as primary and secondary leptons.  
Leptons directly produced in the decay ($b \rightarrow l X$) are called
primary leptons, and those indirectly produced ($b \rightarrow c X \rightarrow
l Y$) are called secondary leptons.  
Primary leptons have a harder energy spectrum than secondary leptons.  
In the case of decays to $e^+$ or $e^-$, a $b$ quark mainly produces primary
electrons and secondary positrons although 
it can also produce a smaller number of
primary positrons due to $B^0-\bar B^0$ mixing.  The branching ratios of
the necessary $b$ quark decays are 9.30\% to primary electrons, 2.07\% to
secondary electrons, 
1.25\% to primary positrons, and 7.36\% to secondary positrons, respectively.  
The branching ratios and energy spectra from JETSET7.4, consistent with
measurements\cite{bdecay}, are almost identical for muons and electrons.

Like dileptons from charm decays, dileptons from bottom decays can come
from correlated and uncorrelated decays of bottom pairs.  
However, uncorrelated pairs from bottom decays are negligible since 
the average number of bottom quarks per central Au+Au event at 
RHIC is much less than 1. The total number of dielectrons from a 
$b\bar b$ decay can be readily estimated to be 0.020.  
Another important source of dileptons from bottom decays is
the decay of a single bottom ($b \rightarrow c l_1 X \rightarrow l_1 l_2 Y$),
unlike the case of charm decays.  
The branching ratio for a $b$ quark to a dielectron is 0.906\% from JETSET7.4,
therefore this source gives 0.018 dielectrons,
comparable to the yield from correlated decays of $b \bar b$ pairs and
thus must be included.  

\begin{figure}[h]
\setlength{\epsfxsize=\textwidth}
\setlength{\epsfysize=0.6\textheight}
\centerline{\epsffile{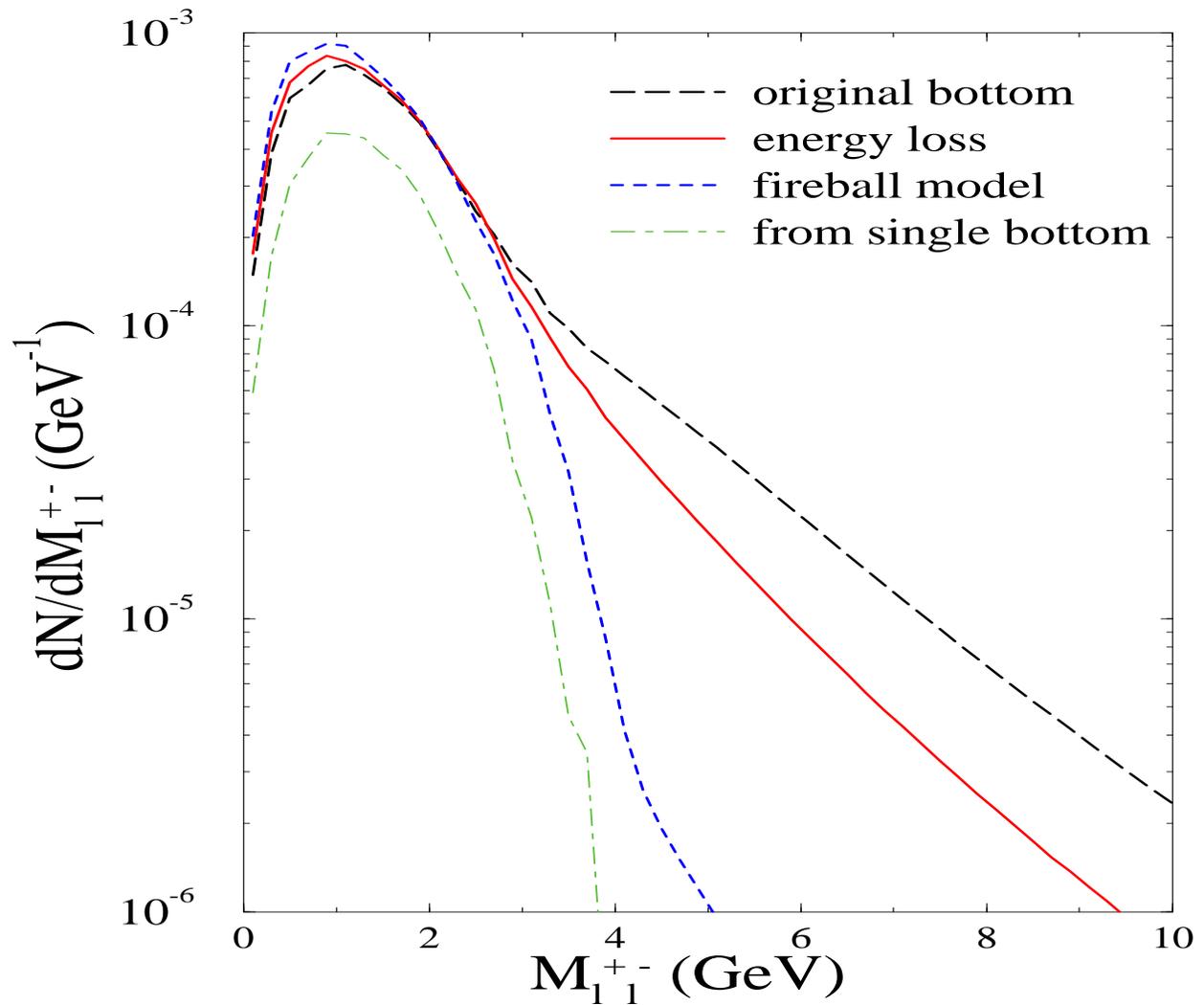}}
\vspace{1cm}
\caption{
Invariant mass spectrum of dileptons from bottom decays.  
The dot-dashed curve is the contribution from the decay of single 
$b$ and $\bar b$ quarks.  
}
\label{fig_bottomll}
\end{figure}

Fig.~\ref{fig_bottomll} shows the invariant mass spectrum of dileptons 
from bottom decays without any kinematic cuts. The low invariant mass 
part of the spectrum is dominated by dileptons from single bottom 
quarks and $b \bar b$ pairs with zero relative momentum. 
The dot-dashed curve comes from single bottom 
decays and thus does not depend on the energy loss.  
The suppression of high invariant mass pairs 
from bottom decays are similar to the charm case shown in 
Fig.~\ref{fig_charmll}.  However, the suppression 
due to the energy loss begins at larger dilepton invariant mass for
bottom decays.

Comparing the solid and long-dashed curves in Figs.~\ref{fig_charmll} and
~\ref{fig_bottomll}, one might suppose that the effect of energy loss on
dileptons from heavy quark decays is only a factor of 2. However, we must 
emphasize that this impression is misleading since this 
spectrum is integrated over all other kinematic variables,  
including rapidities of the single quarks.  
In our model, while a heavy quark loses transverse momentum its rapidity is
essentially unchanged due to the longitudinal flow. 
Because of the rapidity gap between the $Q$ and $\bar Q$, a heavy quark pair as
well as the resulting lepton pair can still have a sizable invariant 
mass after the energy loss.  
However, if one selects leptons with a transverse momenta cut, 
the effect of energy loss becomes much more dramatic, as we will show in the
next section where we consider the finite rapidity coverage and single lepton
momentum cuts of the PHENIX detector. 

\section{Dileptons from Heavy Quark Decays as Seen by PHENIX}
\label{sec-phenix}

The PHENIX detector at RHIC is designed to measure electromagnetic
signals of dense matter. It is therefore well suited to carry 
out single lepton ($e,\mu$) and dilepton ($ee, e\mu, \mu\mu$) 
measurements.  In this section we calculate the dilepton yields 
within the designed PHENIX detector acceptance.  

The PHENIX detector \cite{cdr} has two electron arms and two muon arms.
The electron arms cover the central region with electron
pseudo-rapidity $-0.35 \leq \eta _e \leq 0.35$, and azimuthal angle $\pm
(22.5^\circ \leq \phi_e \leq 112.5^\circ )$.  The muon arms cover the 
forward and backward regions with polar angle $\pm (10^ \circ \leq
\theta_\mu \leq
35^\circ)$, corresponding to the pseudo-rapidity interval $\pm (1.15 \leq
\eta_\mu \leq 2.44)$, along with
almost the entire azimuthal angle coverage.  
We take $E_e > 1$ GeV and $E_\mu > 2$ GeV\cite{cdr} to
reduce the lepton backgrounds from random hadronic decays such as pions.

\begin{figure}[h]
\setlength{\epsfxsize=\textwidth}
\setlength{\epsfysize=0.6\textheight}
\centerline{\epsffile{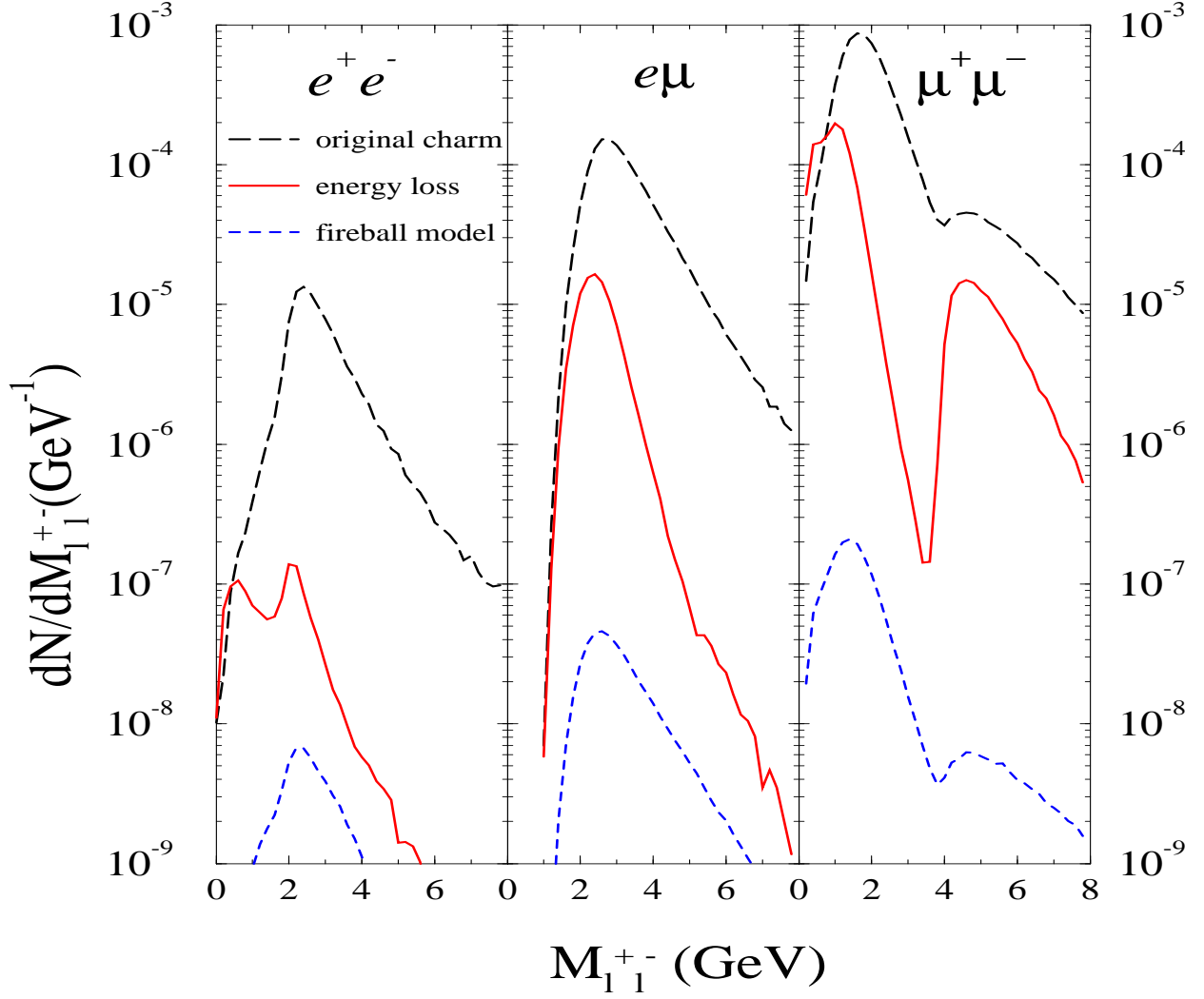}}
\vspace{1cm}
\caption{
Invariant mass spectra of the
three dilepton channels, $ee, e\mu$, and $\mu\mu$,  
from correlated charm pair decays within the  PHENIX acceptance.  
}
\label{fig_charm3M}
\end{figure}

Fig.~\ref{fig_charm3M} shows the invariant mass distribution of three types of
dileptons from open charm decays within the acceptance of PHENIX.  
The $e\mu$ spectrum includes both $e^+\mu^-$ and $e^-\mu^+$.  
From the comparison of our energy loss results with the initial distributions, 
we note that the three dilepton yields have strikingly different suppression
factors.  
In the peaks of the spectra, our calculation shows that the $ee$ yield is
suppressed roughly by 100, the $e\mu$ yield by 10, and the $\mu\mu$ yield by
only 4.  
The static fireball model, however, gives a suppression of about $3000$ 
for all three dilepton channels.

The large difference between the suppression patterns due to energy loss of the
three types of dileptons is a result of our model of the longitudinal flow
combined with the PHENIX rapidity coverage and the single lepton energy cut.  
From the PHENIX detector geometry, the accepted electrons 
have nearly zero rapidity. Therefore the energy cut of 1 GeV 
for single electrons is actually a 1 GeV $p_\perp$ cut.
However, the accepted muons typically have a rapidity of 2 so that 
the 2 GeV energy cut corresponds to a $p_\perp$ cut of only 0.5 GeV.  
As seen in Figs.~\ref{fig_pt_c} and \ref{fig_y_c}, in our model
of longitudinal flow, energy loss only suppresses high $p_\perp$ charm quarks 
without strongly affecting the rapidity spectrum.  
Consequently, high $p_\perp$ leptons from semileptonic charm decays
are strongly suppressed while low $p_\perp$ leptons may be 
slightly enhanced. Therefore the accepted electrons from charm 
decays in PHENIX are more strongly suppressed than the accepted muons. 

\begin{figure}[h]
\setlength{\epsfxsize=\textwidth}
\setlength{\epsfysize=0.6\textheight}
\centerline{\epsffile{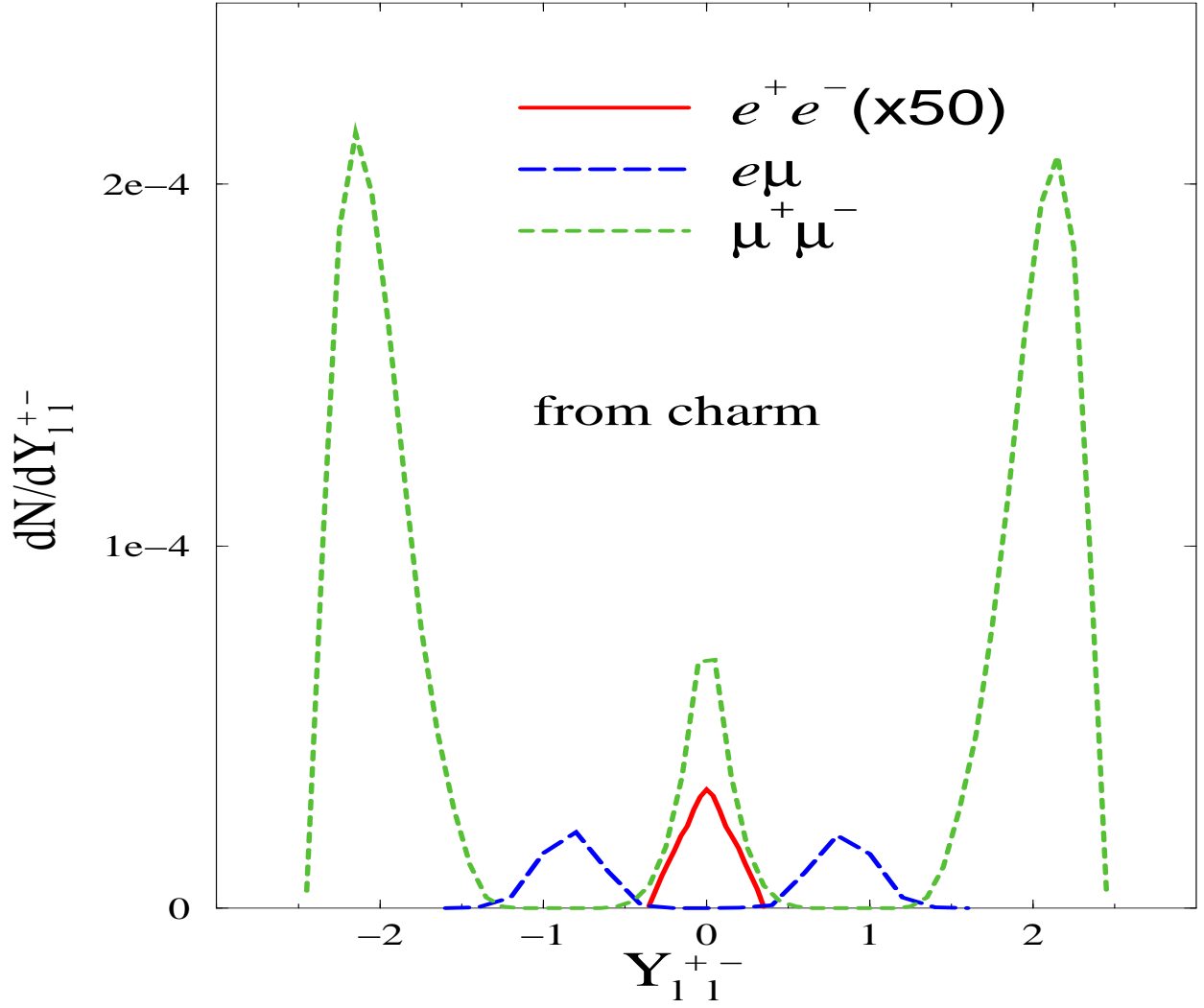}}
\vspace{1cm}
\caption{
Rapidity spectra of the three dilepton channels, $ee, e\mu$, and $\mu\mu$, 
from correlated charm pair decays within the PHENIX acceptance.  
The solid curve representing dielectrons is scaled up by a factor of 50.  
The long-dashed curve represents opposite-sign $e\mu$ pairs.
}
\label{fig_charm3Y}
\end{figure}

To demonstrate the acceptance of the PHENIX detector, in 
Fig.~\ref{fig_charm3Y} we show the rapidity distribution of the three types of
dileptons from charm quark decays as seen by PHENIX.  The $ee$ pairs are
centered  around $Y_{l^+l^-} \sim 0$, while the $e\mu$ acceptance covers 
pair rapidity around $-$1 and 1, 
and the $\mu\mu$ pairs are found with $Y_{l^+l^-} \sim$ $-$2, 0 and 2. 
The two-peak nature of the dimuon 
spectrum in Fig.~\ref{fig_charm3M} is a result of the muon rapidity coverage.  
The low-mass peak comes from two muons in the
same arm while the high-mass peak comes from pairs with one muon in 
each arm.  Similarly, the two-peak nature of the dielectron 
spectrum in Fig.~\ref{fig_charm3M} is a result of the two azimuthal 
electron coverages in PHENIX.

The strong difference in the suppression of the three 
dilepton spectra from charm quark decay in PHENIX, as seen in
Fig.~\ref{fig_charm3M}, suggests that we should carefully consider how to
measure open charm with the PHENIX detector.  In a naive scenario, one
would expect that the $e \mu$ coincidence measurement can reveal 
the open charm cross section so that the dielectron and 
dimuon open charm yields can be extrapolated just by considering 
the different rapidity coverages.  
However, energy loss complicates the acceptances for the three dilepton
channels from open charm decays, making the proposed extrapolation  
difficult.  As we will see later in this section, this 
extrapolation becomes even more complicated when bottom decays are included.

The large suppression of dileptons from open charm decays implies that we 
should also consider dileptons from bottom decays.  
Due to their larger mass, bottom quarks produce more energetic 
leptons than do charm quarks, thus dileptons from bottom decays are 
expected to be less suppressed than those from charm decays and 
therefore may become important.  

\begin{figure}[h]
\setlength{\epsfxsize=\textwidth}
\setlength{\epsfysize=0.6\textheight}
\centerline{\epsffile{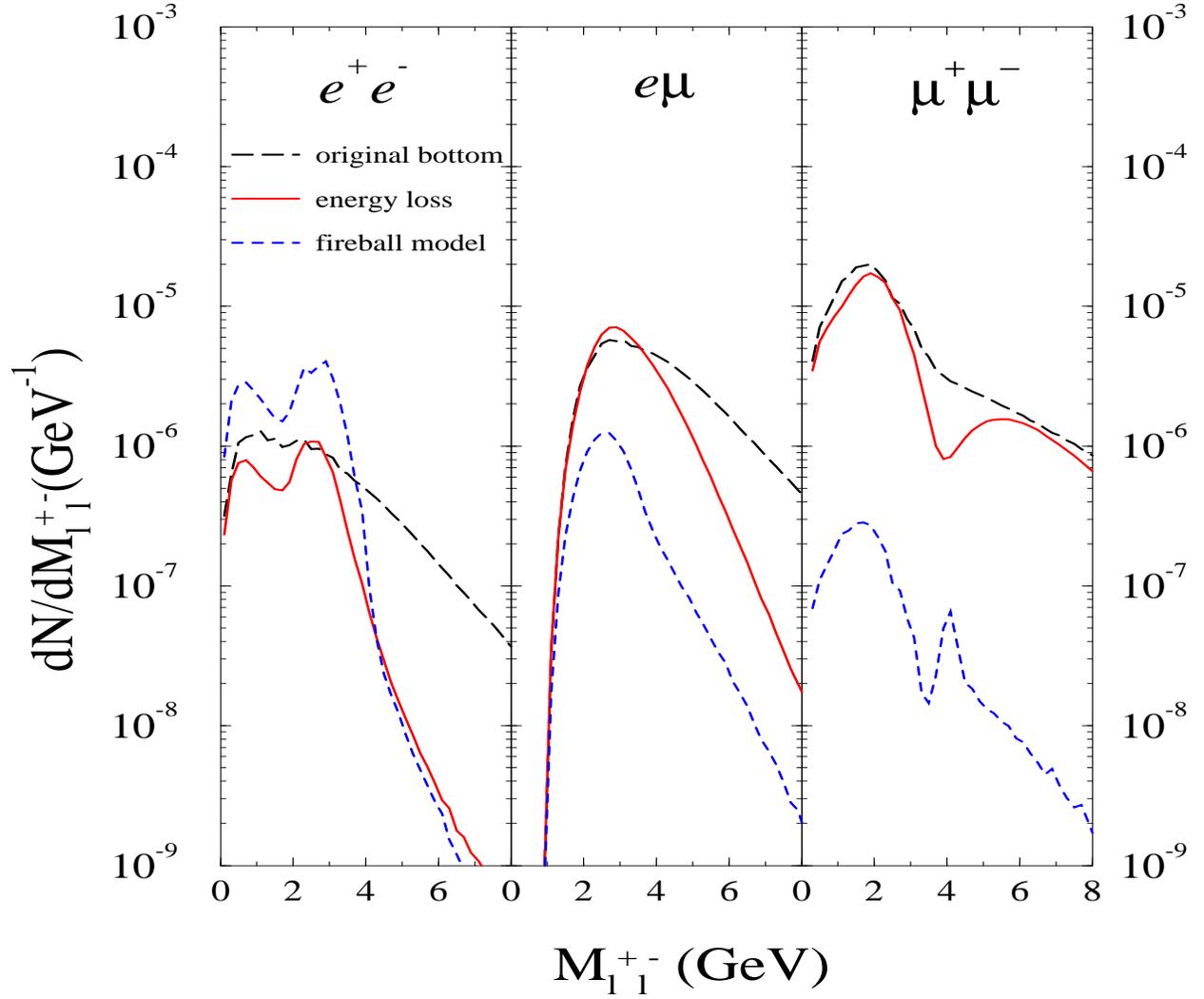}}
\vspace{1cm}
\caption{
Invariant mass spectra of the 
three dilepton channels, $ee, e\mu$, and $\mu\mu$,  
from bottom decays within the PHENIX acceptance.  
}
\label{fig_bottom3M}
\end{figure}

\begin{figure}[h]
\setlength{\epsfxsize=\textwidth}
\setlength{\epsfysize=0.6\textheight}
\centerline{\epsffile{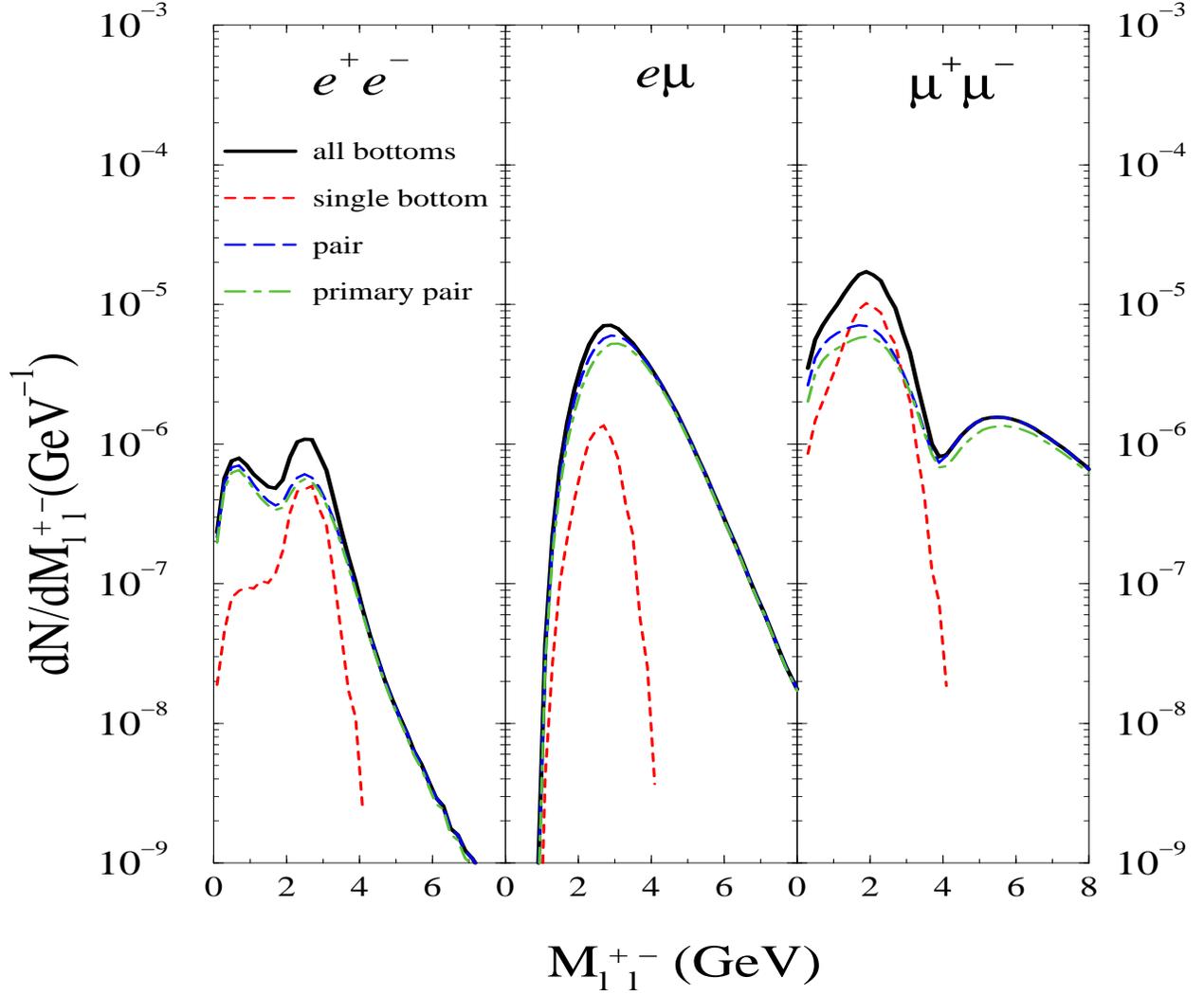}}
\vspace{1cm}
\caption{
The three dilepton channels from bottom decays in our model within the PHENIX
acceptance.  
The short-dashed curves and the long-dashed curves represent dileptons from
single quark decays and bottom pair decays, respectively.
The solid curves represent the sum of these two curves.  
The dot-dashed curves represent dileptons from bottom pair decays where both
leptons are primary leptons.  
}
\label{fig_bottomsource}
\end{figure}

Fig.~\ref{fig_bottom3M} shows the dilepton yields from 
bottom decays when the PHENIX geometry and energy cuts are applied.  
Although there is significant suppression due to energy loss at high
invariant mass, the peaks of the spectra are not strongly suppressed.  
In the static fireball model dielectrons are even enhanced because 
there are more stopped bottom quarks at zero rapidity and a bottom quark at
rest can still produce an electron energetic enough to pass the 1 GeV cut.
The dilepton sources from bottom decays in our energy loss model, shown by the
solid curves in Fig.~\ref{fig_bottom3M}, are categorized in
Fig.~\ref{fig_bottomsource}.  
Dileptons from single $b$ or $\bar b$
decays can only be significant at low invariant mass.  At high invariant mass, 
the dileptons all come from decays of $b \bar b$ pairs.  
Within the PHENIX acceptance, dileptons from bottom pair decays are 
dominated by dileptons consisting of two primary leptons because 
primary leptons are more energetic and therefore favored after the energy cuts.

\begin{figure}[h]
\setlength{\epsfxsize=\textwidth}
\setlength{\epsfysize=0.6\textheight}
\centerline{\epsffile{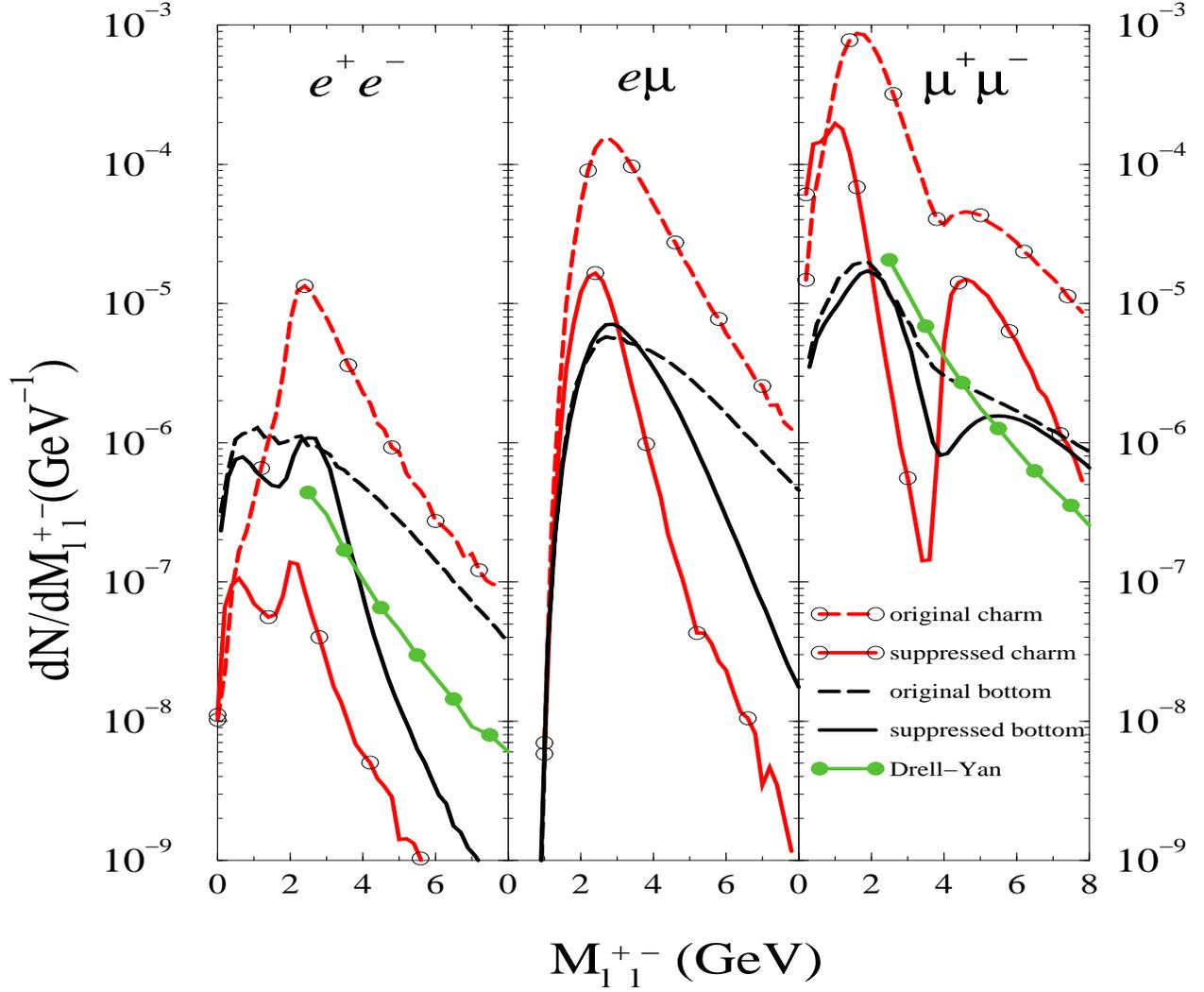}}
\vspace{1cm}
\caption{
Invariant mass spectra of the three dilepton channels from correlated
bottom and charm decays within the PHENIX acceptance.  
The $ee$ and $\mu\mu$ channels are compared to the Drell-Yan yields.
}
\label{fig_hvq}
\end{figure}

We compare dileptons from bottom decays with those from correlated 
charm decays in Fig.~\ref{fig_hvq}.  We also calculate the
Drell-Yan yields of dielectrons and dimuons from the HIJING program and 
plot them in the same figure.  
Without energy loss, dileptons from charm dominate 
those from bottom decays as well as the Drell-Yan yield in the invariant mass
region below 8 GeV.  
With an energy loss rate of $-$2 GeV/fm, dielectrons from $b \bar b$ decays
dominate those from charm and the Drell-Yan rate dominates both at high
invariant mass.  
This reduction of continuum dileptons from open charm and bottom decays
certainly provides a better opportunity to observe possible thermal dileptons
from the quark-gluon plasma.   

With an energy loss rate of $-$2 GeV/fm, the $e\mu$ yield is dominated by
bottom decays at high invariant mass.
The Drell-Yan process does not contribute to the $e\mu$ channel.   
Thus high-mass $e\mu$ pairs can be used for bottom observation if this energy
loss rate will be reached.

The Drell-Yan dimuon yield lies below the second dimuon peak from the
suppressed charm quarks at invariant masses above 4 GeV.  
Moreover, most of the Drell-Yan dimuons are in the same
muon arm due to the strong rapidity correlation.  
The second dimuon peak mainly comes from the pairs with two muons in
opposite muon arms.  
Therefore it is possible to suppress the Drell-Yan dimuons by imposing
large-angle cuts and use high-mass dimuons for charm observation, at
least in the rapidity region around 0.   

\begin{figure}[h]
\setlength{\epsfxsize=\textwidth}
\setlength{\epsfysize=0.6\textheight}
\centerline{\epsffile{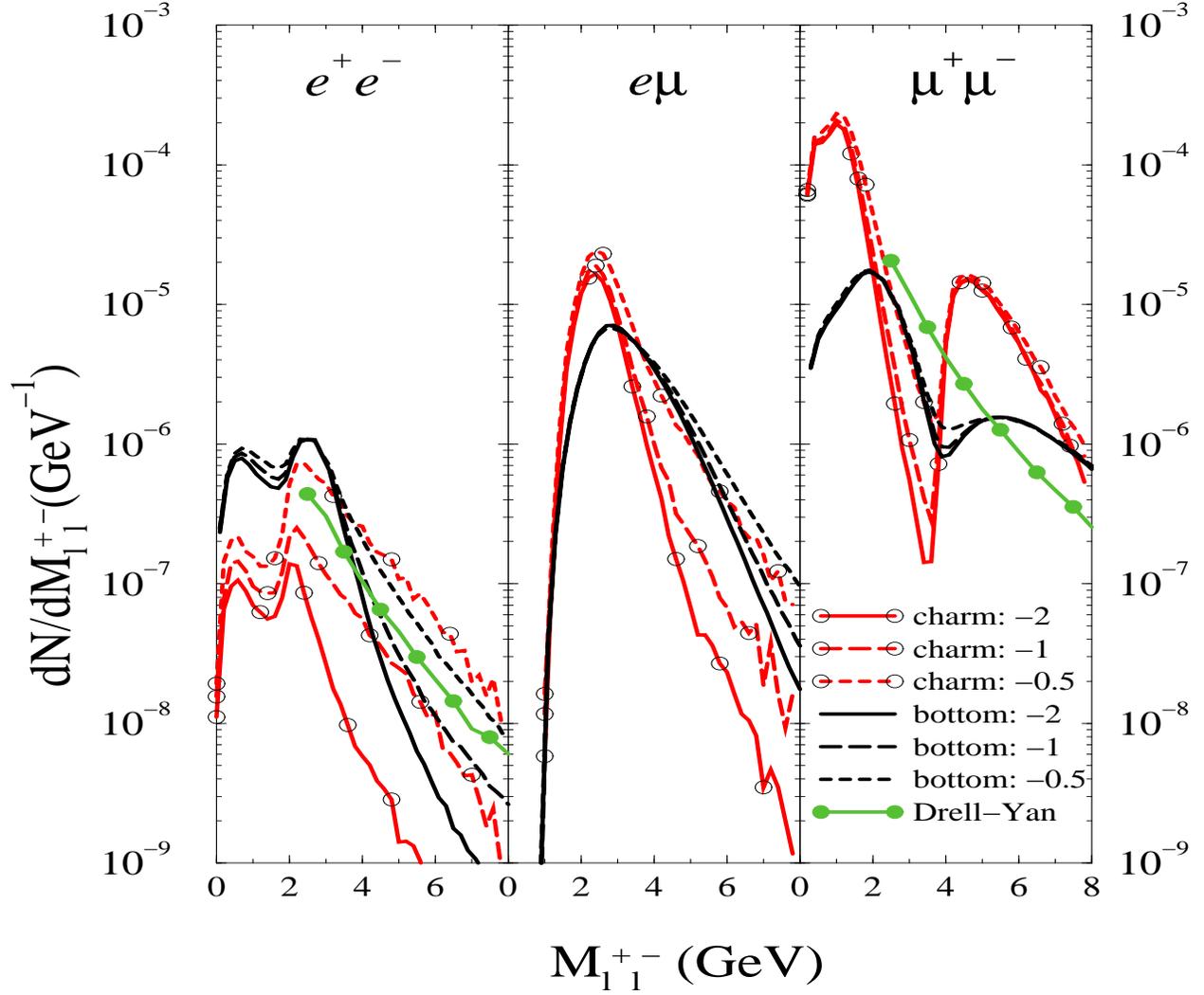}}
\vspace{1cm}
\caption{
Dileptons from charm and bottom decays within the PHENIX acceptance 
with different values of $dE/dx$ ($-$2, $-$1, and $-$0.5 GeV/fm).  
}
\label{fig_hvq3}
\end{figure}

To study the sensitivity of the dilepton spectra from heavy quark
decays to the energy loss rate, in Fig.~\ref{fig_hvq3} we compare calculations 
with three values of $dE/dx$, $-2$ GeV/fm, $-1$ GeV/fm, and $-0.5$ GeV/fm.
Within the PHENIX acceptance, high invariant mass dielectrons are the most
sensitive to the energy loss.  
This is expected because high invariant mass dielectrons mainly come from
decays of two energetic heavy quarks.  
As a result of the corona effect, the suppression of high invariant mass
dielectrons will saturate at large $-dE/dx$ for a fixed $\lambda$.
However, low invariant mass dielectrons have a significant
contribution from decays of two thermalized heavy quarks. 
Therefore they are not sensitive to the energy loss and the suppression of low
invariant mass dielectrons will saturate at large $-dE/dx$.  

With a smaller energy loss rate of $-$0.5 GeV/fm, dielectron yields from charm
and bottom decays are much less suppressed, and become larger than the
Drell-Yan yield. 
This sensitivity of the dielectron yields to the energy loss can be used to
probe the property of the dense medium.  Dileptons from bottom decays are less
sensitive to the energy loss than those from charm decays.  As a result, the
$e\mu$ yield from bottom decays becomes less dominant relative to that from
charm decays when the parameter $dE/dx$ becomes smaller.  Since 
bottom quarks may lose less energy than charm quarks, the $e\mu$ channel may
still be dominated by bottom decays and thus provide a measure of bottom
production. 
Dimuons are the least sensitive to the energy loss rate.  Dimuons with 
invariant mass above 4 GeV are dominated by charm decays, and are therefore
useful for charm detection.

Energy loss is described by two parameters in our model, $dE/dx$ 
and $\lambda$, where $dE/dx$ describes the average energy loss rate of a 
fast parton and the mean free path, $\lambda$, determines the frequency of 
scatterings.  A larger $\lambda$ results in a larger 
probability for the fast parton to escape without energy loss, similar to a
smaller $dE/dx$. This is responsible for 
the difference in the large-$p_\perp$ suppression between 
our model and the static fireball model. If we take the limit $\lambda=0$,
the models should agree at large $p_\perp$.  One can also probe these
two parameters simultaneously as in the study of jet quenching in
$\gamma$-jet events \cite{whs}. However, it is not yet clear 
how to disentangle these two effects.

\section{Single Leptons from Heavy Quark Decays}
\label{sec-single}

Single leptons from charm decay have been suggested as an indirect measure
of the charm production cross section \cite{tann}. This is possible if
the background leptons from random decays of hadrons such as pions and kaons
can be well understood. 
  
\begin{figure}[h]
\setlength{\epsfxsize=\textwidth}
\setlength{\epsfysize=0.6\textheight}
\centerline{\epsffile{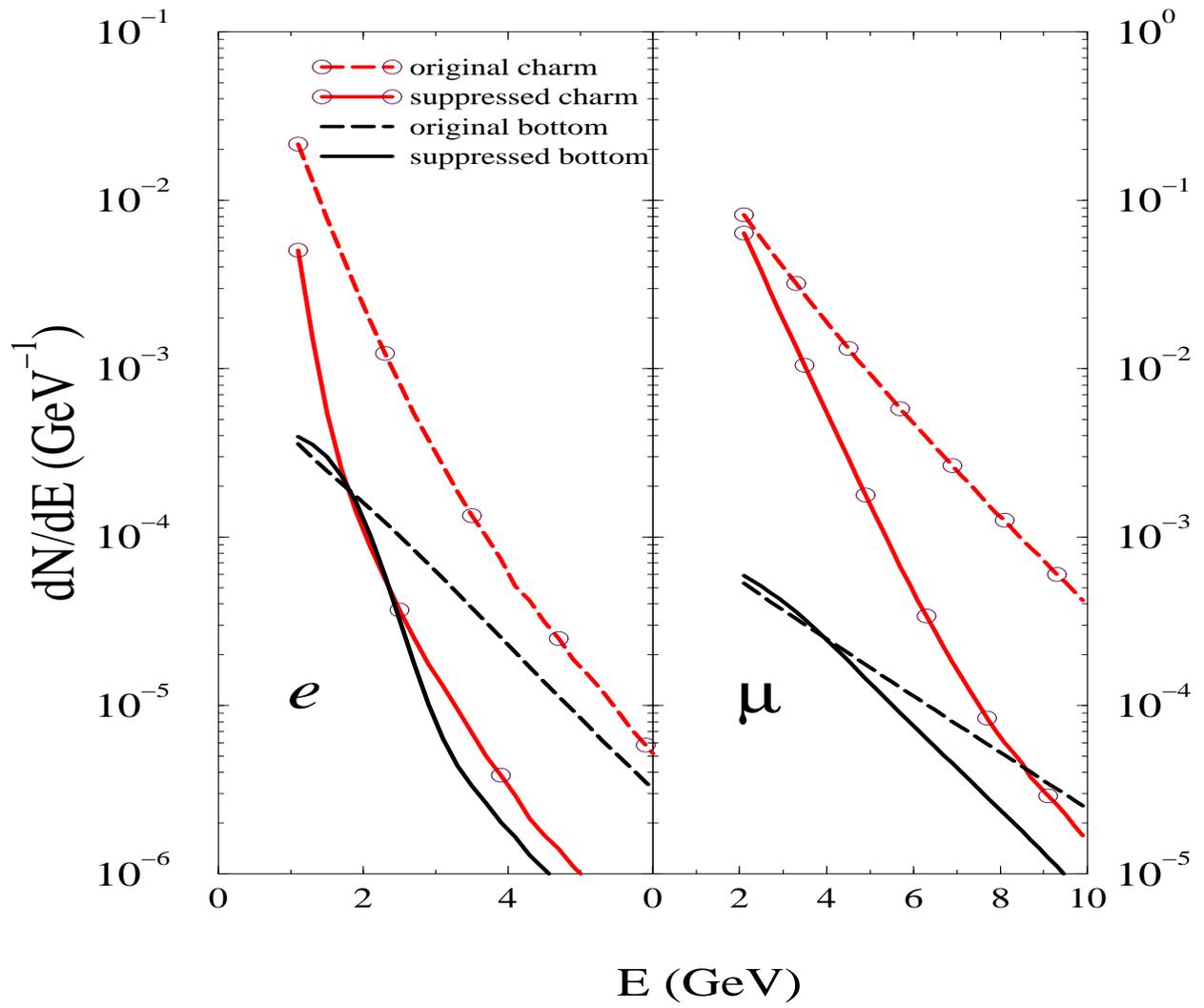}}
\vspace{1cm}
\caption{
Energy spectrum of single electrons and muons from charm and bottom
decays within the PHENIX acceptance.  
}
\label{fig_single}
\end{figure}

\begin{figure}[h]
\setlength{\epsfxsize=\textwidth}
\setlength{\epsfysize=0.6\textheight}
\centerline{\epsffile{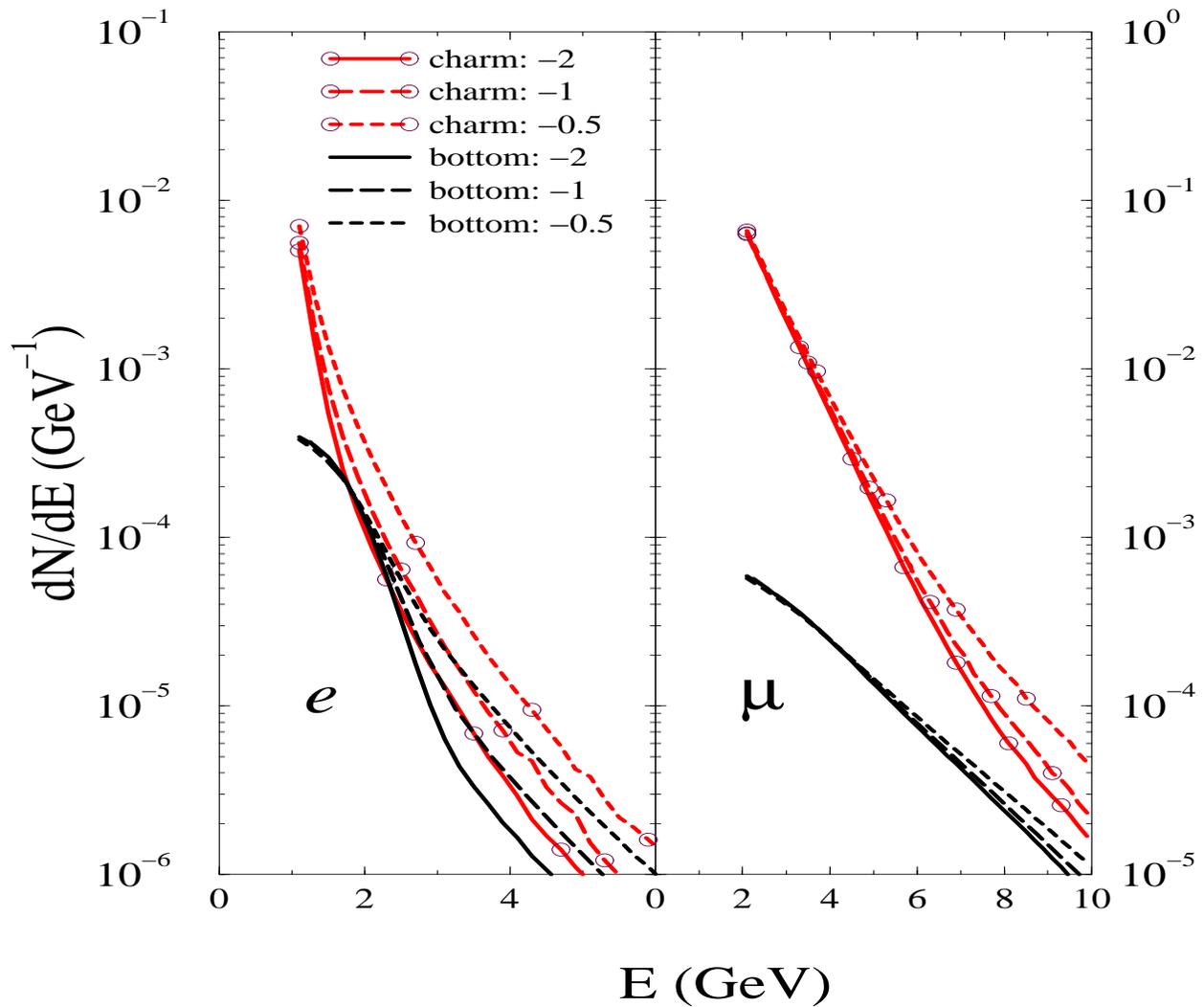}}
\vspace{1cm}
\caption{
Single electrons and muons from charm and bottom decays within the PHENIX
acceptance with different values of $dE/dx$ ($-$2, $-$1, and $-$0.5 GeV/fm). 
}
\label{fig_single3}
\end{figure}

We show the effect of energy loss on single electrons and muons within the
PHENIX acceptance in Fig.~\ref{fig_single}.  
With energy loss, the electron yield from charm decays is close to that from
bottom decays at energies above 2 GeV.  
Muons from charm decays are less dominant compared to those from bottom
decays, though it seems possible to use single muons to measure the open
charm cross section.  
Calculations with three values of $dE/dx$, $-2$ GeV/fm, $-1$ GeV/fm, and $-0.5$
GeV/fm, are compared in Fig.~\ref{fig_single3}.  
Single leptons are not as sensitive to the magnitude of $dE/dx$ as the dilepton
mass spectra.  

Single leptons can be categorized as those from thermalized
heavy quarks and those from heavy quarks energetic enough to
escape after energy loss. 
The former mainly reflects the effective
thermalization temperature while the latter can provide  
us with information on the energy loss.  
Single electrons with energies greater than $2-3$ GeV 
are mainly from energetic heavy quarks and thus are more sensitive to the
energy loss.  Muons from energetic heavy quarks start to dominate the spectrum
at a higher energy.  

\section{Summary and Discussion}
\label{sec-summary}

Dileptons from open charm and bottom decays have been calculated for central
Au+Au events at RHIC including the effect of energy loss on heavy quarks
in dense matter.  We assumed a Bjorken-type longitudinal expansion 
where heavy quarks move with the flow and thus basically do not change their
rapidities throughout the evolution of the system.
Heavy quarks, however, lose transverse momentum when traveling through
the dense matter.  
Compared to the case without energy loss, heavy quarks at large transverse
momentum are greatly suppressed.
As a result, dileptons from heavy quark decays are suppressed at
high invariant mass.  

With our model of the energy loss, the three types of dileptons ($ee, e\mu,
\mu\mu$) from heavy quark decays within the PHENIX acceptance are suppressed 
very differently because of the energy cuts and the different rapidity
coverages.  
Thus it is difficult to extrapolate the dielectron and dimuon
yields from charm decays based on the $e\mu$ coincidence measurement.  
Within the PHENIX acceptance, dielectrons are the most sensitive to the 
energy loss,   
and dielectrons from heavy quark decays become comparable or even lower than
the Drell-Yan yield at high invariant mass.
Therefore, they are the best observables to study the
energy loss if the heavy quark spectrum cannot be measured
directly via traditional tracking techniques. 
Contrary to the case without energy loss, at high invariant mass the
$e\mu$ yield may very well be dominated by bottom decays 
instead of charm decays.  
In this case, $e\mu$ coincidence in $A+A$ collisions can no longer measure the
charm contribution, but could be used for bottom observation.  
Dimuons above 4 GeV, which are well above the Drell-Yan yield even with a large
energy loss rate, can be used for charm observation.  

There are a number of uncertainties in our treatment of the energy loss.  
There are no calculations of the radiative energy loss rate for massive quarks
in medium, therefore $dE/dx$ is a parameter in our model.  
We have neglected the relative formation times.  However, the longitudinal 
velocity of heavy quarks and the dense matter fluid could be mismatched and
this will quantitatively affect the results. 
The rapidity distribution of the 
final heavy quarks depends sensitively on the flow pattern.  
If the different suppressions of the three dilepton channels ($ee,
e\mu,\mu\mu$) can be established, it will provide important information about
the fluid dynamics of the dense medium.
We have not considered any change in the energy loss which could
be caused by the expansion of 
the system and the subsequent decrease in the energy density.
Transverse flow is not included in our ideal Bjorken picture
which could quantitatively change the low
$p_\perp$ lepton and low invariant mass dilepton yields in our
calculations.  
However, the qualitative features of our results, 
such as the suppression of high $p_\perp$ leptons and high
invariant mass dileptons as well as the different suppression patterns for the
dilepton channels at PHENIX are not likely to change.  

To determine whether single leptons or dileptons from heavy quark
decays can indeed probe the energy loss, the most important factor is the
background from random hadronic decays.  
Though we have not done an analysis of the hadronic decays in this paper, it
needs further study, particularly since high $p_\perp$ pions will also 
experience quenching effects and therefore be suppressed in high-energy
heavy-ion collisions as well.  

Acknowledgments: We thank M. Gyulassy for the inspiration and discussions.  
We thank E. Shuryak for discussions related to his earlier study on this
subject.  We also thank T. LeCompte and W. Zajc for helpful discussions.  

\pagebreak

\pagebreak

\pagebreak

\pagebreak

\pagebreak

\pagebreak

\pagebreak

\pagebreak

\pagebreak

\pagebreak

\pagebreak

\pagebreak

\pagebreak
{}

\end{document}